\documentclass{article}

\newcommand{\be}{\begin{equation}}    
\newcommand{\ee}{\end{equation}}
\newcommand{\ba}{\begin{eqnarray}}
\newcommand{\ea}{\end{eqnarray}}
\newcommand{\sizedef}{
        \headheight=0pt                               
 	  \topmargin=-1.5cm \headsep=1.5cm              
        \oddsidemargin=-0.5cm \evensidemargin=-0.5cm  
        \textheight=22truecm \textwidth=16.5truecm    
 	  \setlength{\columnsep}{20pt}                  
  }
 \usepackage{amsmath,amssymb}
\sizedef

\begin{document}
\bibliographystyle{prsty}

\title{Configurational invariants of Hamiltonian systems}

\author{ 
Giuseppe Pucacco\thanks{e-mail: pucacco@roma2.infn.it} \\ 
Dipartimento di Fisica -- Universit\`a di Roma ``Tor Vergata" \\ 
INFN -- Sezione di Roma II\\
\\
Kjell Rosquist\thanks{e-mail: kr@physto.se} \\
Department of Physics -- Stockholm University} 
\date{}
\maketitle 
\vspace{2cm}
 
\begin{abstract}

\noindent
In this paper we explore the general conditions in order that a 2-dimensional natural Hamiltonian 
system possess a second invariant which is a polynomial in the momenta and is therefore Liouville
integrable. We examine the possibility that the invariant is preserved by the Hamiltonian flow on a
given energy hypersurface only (weak integrability) and derive the additional requirement necessary
to have conservation at arbitrary energy (strong integrability). Using null complex coordinates, we show that the leading order coefficient of the polynomial is an arbitrary holomorphic function in
the case of weak integrability and a polynomial in the coordinates in the strongly integrable one. We review the results obtained so far with strong invariants up to degree four and provide some new examples of weakly integrable systems with linear and quadratic invariants. 
\end{abstract}

\vskip2cm

\noindent
Published as: {\it Journal of Math. Phys.}, {\bf 46}, 052902 (2005)


\section{Introduction}

In 1983 Hall \cite{hall} published a remarkable paper devoted to a theory of
{\it configurational invariants} of classical Hamiltonian systems. The main
virtue of the work consisted in an elegant and powerful technique to solve the
equations for the existence of a second polynomial invariant of arbitrary
degree in the momenta for 2-dimensional Hamiltonian systems. As a result of this
approach, Hall was able to get some new examples of integrable systems admitting a
second invariant of degree four in the momenta.

Unfortunately, the paper was flawed by a definitely wrong statement and also by many inaccurate
arguments and deductions. In fact, Hall purported to remedy
supposed oversights in previous works on the search for the second
invariant, in particular criticizing the classical account by Whittaker
\cite{whit}. In Hall's view, Whittaker's (and all others' since then) treatment provides only sufficient conditions for the existence of a second invariant (linear and quadratic in the specific instance), whereas, due to overlooking the link established on phase-space variables by energy conservation, it was not able to find all possible solutions. Actually, as it is, this statement is wrong: in looking for a {\it strong} second invariant, namely a phase-space function which commutes with the Hamiltonian function, Whittaker's approach is indeed correct and leads to {\it necessary and sufficient} conditions for its existence. This point was already stressed by Sarlet et al.\ \cite{SLC85a} in their criticism to Hall's paper. 

Moreover, Hall's discussion contained a confusion between the concept of
configurational (or {\it weak}) invariant, as a function which exactly commutes with the Hamiltonian only on a subset (possibly one) of the energy hypersurfaces (see again \cite{SLC85a} and, e.g.\ \cite{BP1}), and the notion of what we may call {\it formal integral} as it emerges in the analysis of regular portions of the phase space of generic non-integrable systems (see e.g.\ \cite{moser}). In particular, the approximate invariants obtained by Hall as a result of his perturbative approach have little to do with those that can be obtained by truncating a normal form expansion.

Nonetheless, in spite of all the above shortcomings, the form in which
the problem has been set by Hall and the approach followed for its partial solution deserve attention, since they can still be very useful. Already Hietarinta \cite{hiet}, in his account of the direct methods for the search of the second invariant, provides a review of all the known systems admitting one
or more configurational invariants and works out again the integrability
conditions found by Hall for the existence of weak invariants up to degree
four. However, in Hietarinta's review, the two settings of weak and strong integrability are still kept well separated. More recently, in a series of works about a unified approach to treat both kinds of invariants \cite{geom,max,kprs,pr05}, the same integrability conditions have been obtained,  {\it supplemented by the additional constraint imposed by strong integrability}. This last step is essential for a neat distinction between the two notions of weak and strong integrability. In these works, this step is a straightforward consequence of the geometric approach in which the existence of the second invariant is addressed by studying the corresponding Killing tensor equations for the Jacobi metric. This metric depends on the mechanical energy as a parameter. Therefore, any tensor object on the corresponding manifold in general depends on the energy. With due care about formal relations between the two approaches, the geometric approach and Hall's approach are equivalent. The essential remark is that, to identify the cases of strong integrability, it is sufficient that in the final results concerning the existence of the invariant, a subset can be isolated which is independent of the energy parameter. In this framework, in \cite{geom}, quadratic invariants at arbitrary and fixed energy for 2-dimensional Hamiltonian systems were treated in a unified way, whereas in \cite{max,kprs} the existence of respectively {\em cubic} and {\em quartic} invariants was discussed accompanied with the discovery of some new examples not given in earlier works (see e.g.\ \cite{dgr,gdr1,gdr2,eva}). In \cite{pr05} the case of Hamiltonian systems with vector potentials is treated, extending previous investigations \cite{dgrw,msw,bw,y1,y2,bcr}.

The aim of the present paper is to discuss the techniques for solving the equations for a second invariant having a momentum dependence of arbitrary polynomial degree. The analysis is based on a combination of both the above-mentioned approaches, with particular attention dedicated to the conformal transformations used to simplify the equations. The treatment is in general effective for both classes of invariants, configurational (or weak) and strong. After that, we impose the additional condition needed to isolate the class of strongly integrable systems, obtaining the general form of the simplifying family of transformations for each degree of the strong invariant looked for. This in turn implies determining the leading order terms in the invariant itself. An alternative route to this result in the geometric approach is based on the invariance, under conformal transformations, of the conformal part of the Killing tensor. Moreover, a general review of the results obtained so far is provided.

The plan of the paper is as follows: in Section 2 we compare the notion of weak and strong invariants and, working out in detail the quadratic case, we correct Hall's misunderstanding, providing the constraint to be satisfied in order to get strong integrability as a restriction of weak integrability; in Section 3 we recall time reparametrization of the null Hamiltonian and
complex (``null") coordinates used to simplify the direct approach; in Section 4 these tools are exploited to set the general approach to find polynomial invariants of arbitrary degree; in Section 5 we recall the main results concerning invariants of degree up to four; in section 6 we present our conclusions and the prospects for future works.

\section{Configurational invariants versus strong invariants}

As it is well known (see, e.g.\ \cite{arnold:mechanics}), to grant the complete
integrability of an $N$-dimensional Hamiltonian system, it is necessary and sufficient to find $N$ independent integrals of motion or {\it invariants}, for short. In the following, we will limit ourselves to the simplest case of a conservative Hamiltonian system in two dimensions ($N=2$). Since the  Hamiltonian itself
\begin{equation}\label{H} 
H =  {\scriptstyle \frac12}  (p_x^2 + p_y^2) + V (x, y), 
\end{equation}
is a conserved function, it is enough to find just a second independent invariant.

In his search for polynomial invariants, Hall
\cite{hall} exploits what is usually called the ``direct approach" of Darboux
\cite{darboux:inv} and Whittaker
\cite{whit}. As an ``educated guess" we may start with the working hypothesis of an invariant with a structure analogous to that of the Hamiltonian, namely a second degree polynomial in the momenta.\footnote{Since all the work is made in a Hamiltonian context, contrary to the original treatment by Hall, we use canonical phase-space coordinates. However, we have tried to stay as close as possible to his treatment as refers to the procedure and notations.} Clearly, one could start with a polynomial of arbitrary degree or even with a more general expression. In the core part of the paper we will examine the general case. Now, as an introduction aimed at giving the proper settings, we work out in detail the quadratic case. We may then look for a phase-space function of the form
\begin{equation}\label{I2} 
I_2 (p_x, p_y, x, y) = A p_x^2 + B p_x p_y + C p_y^2 +K, 
\end{equation}
which is preserved along the flow given by eq.(\ref{H}), namely
\begin{equation}\label{poisson}
\{I_2, H \} = 0.
\end{equation}
In eq.(\ref{I2}), $A, B, C$ and $K$ are each functions of $x$ and $y$.
Actually Hall, probably motivated by his interest in accelerator physics, manages to work out the integrability of Hamiltonians including also a vector potential and henceforth terms which are linear in the momenta. This, in turn, suggests the inclusion of analogous terms in the invariant. However, in the standard case of a Hamiltonian invariant under momentum inversion, $p \rightarrow -p$, the equations for the coefficients of even and odd degree terms decouple, greatly reducing the complexity of the system to solve. Therefore, since the main purpose of this paper is to fully clarify the issues of weak and strong integrability, which are unaffected by such generalizations, we prefer to limit ourselves to the standard case of eq.(\ref{H}). The treatment of the vector potential is presented elsewhere \cite{pr05} where new strongly integrable systems with quadratic invariants are presented.

In the direct approach, one inserts the functions (\ref{H},\ref{I2}) into
the Poisson bracket (\ref{poisson}). The resulting polynomial in the momenta, of third degree in the present instance, must be identically vanishing. In view of the independence and arbitrariness of the momentum coordinates, each coefficient of the polynomial must vanish, determining in turn the following system of PDEs in the coordinates: 
\begin{align}\label{QIC}
            A_x &= 0, \\
            A_y + B_x &= 0,\\
            B_y + C_x &= 0, \\
            C_y &= 0,\\
     K_x &= 2 A V_x + B V_y, \\
     K_y &= B V_x + 2 C V_y,
\end{align}
where, in order to compactify expressions, suffixes denote partial differentiation. As mentioned above, Darboux \cite{darboux:inv} addressed the problem and the equation ensuing from the integrability condition for $K$, that is
\begin{equation}\label{darboux}
B (V_{yy} - V_{xx}) +
2 (A - C) V_{xy} + 3 (B_y V_y - B_x V_x) = 0,\end{equation}
is known as {\it Darboux's equation}. Whittaker \cite{whit} also analyzed the problem and, for a complete account of the solution of this system, the standard reference is again the review by Hietarinta \cite{hiet}. 

Equations involving only the leading order terms ($A,B$ and $C$) in the invariant are readily solved:
\begin{align}
\label{LOQIC1}
A &= a y^2 + b y + c,\\
B &= - 2axy - bx - dy - e,\\
\label{LOQIC3}
C &= a x^2 + d x + f.
\end{align}
Exploiting linear transformations of the coordinates, it can be shown \cite{hiet,geom} that these functions can be reduced to four canonical
forms that, inserted into Darboux equation (\ref{darboux}), lead to the complete solution of the problem: there exist four fundamental separating coordinate systems (elliptical, polar, parabolic and Cartesian) in which the potential and the remaining unknown of the invariant, $K$, can be expressed in terms of combinations of arbitrary functions of each of the separating coordinates. Actually, in his reference to Whittaker's result, Hall mentions only the elliptical solution, but, as shown in \cite{geom} (see also \cite{SPT}), using conformal transformations, all the separable cases possess the same structure. Moreover, we remark on the additional possibility of complex potentials, which provides four more cases with complex separating coordinates, which are listed in \cite{hiet} and can also be obtained with the techniques in \cite{geom}.

At this point Hall states that (\cite{hall}, p.93) ``Whittaker's work /.../ was flawed by his failure to recognize that terms in $\dot x^2$ and $\dot y^2$ are not independent for the purpose of setting the coefficients of their various powers to zero /.../ They are related by conservation of energy. Therefore, Whittaker's constraints were over-restrictive; the solutions he found are valid, but others may also exist." On these bases, Hall elaborates a generalized approach that, in his view, provides necessary and sufficient conditions for the existence of the given invariant. Actually, as we will shortly see, Hall fails in recognizing that the possible extension of the family of solutions implies a different status for the additional invariants: they are ``fixed energy" or {\it configurational} invariants. 

Configurational (or ``conditional'' or even ``weak'') integrals hold for {\it a specified particular value of the energy constant} and were firstly investigated by Birkhoff \cite{birkhoff} and considered by Fomenko \cite{fomenko}, Kozlov \cite{kozlov} and others. Weak invariants enjoy weaker properties then their ``nobler" cousins, the {\it strong} invariants, which are constant on {\it every} energy
hypersurface admitted by the dynamics of the system. To clarify this essential
point, let us again work with the quadratic case, concretely introducing Hall's argument. This goes as follows: for every value $E$ of the energy, the
function (\ref{H}) defines the hypersurface
\begin{equation}\label{Ezero} 
   \tfrac12  (p_x^2 + p_y^2) + V (x, y) = E. 
\end{equation}
Let us construct the following linear combinations of the squares of the momenta:
\begin{align}
\Sigma &= p_x^2 + p_y^2, \\
\Delta &= p_x^2 - p_y^2
\end{align}
and observe that, in view of (\ref{Ezero}), we can impose the constraint
\begin{equation}\label{EC}
\Sigma = 2 G(x,y),
\end{equation}
where we have introduced the {\it ``Jacobi" potential\/}\footnote{The origin of this denomination comes from the close relationship between the picture of a system constrained on the fixed energy surface and the geometric picture of a geodesic flow over a Riemannian manifold endowed with a ``Jacobi" metric (see, e.g. \cite{lanczos:mechanics}).}
\begin{equation}\label{G}
G = E - V.
\end{equation}
The quadratic terms in the standard form of the invariant of eq.(\ref{I2}) can now
be written as 
\begin{equation}
A p_x^2 + B p_x p_y + C p_y^2 = 
\frac12 \left[(A+C) \Sigma + (A-C) \Delta + 
 B \sqrt{\Sigma^2 - \Delta^2} \right].
\end{equation}
Exploiting the energy constraint (\ref{EC}) and redefining coefficients, the
quadratic invariant can then be written as
\begin{equation}\label{weakI2} 
I_2 (p_x, p_y, x, y) = 
   \tfrac12  D (p_x^2 - p_y^2) + B p_x p_y + {\widetilde K}, 
\end{equation}
where
\begin{align}
D &= A-C, \label{CK1}\\
{\widetilde K} &= K + (A+C)G. \label{CK2}
\end{align}
Now, we may proceed along the same lines followed above. The commutation relations (\ref{poisson})  
of the weak invariant (\ref{weakI2}) with the Hamiltonian reduce to the system
\begin{align}
             D_y + B_x &= 0, \label{WQIC1}\\
             D_x - B_y &= 0, \label{WQIC2}\\
     {\widetilde K}_x &= B V_y + D V_x - G B_y, \\
     {\widetilde K}_y &= B V_x - D V_y - G B_x.
\end{align}
The integrability condition for ${\widetilde K}$ is now the {\it generalized Darboux equation}
\begin{equation}\label{weakdarboux}
B (V_{yy} - V_{xx}) + 
2 D V_{xy} + 3 (B_y V_y - B_x V_x) - 2 (E - V) D_{xy}= 0,\end{equation}
where we have explicitly pointed out the presence of the energy parameter. In
practice, since the energy of the system is in general an
arbitrary real number, we can write eq.(\ref{weakdarboux}) in the form 
\begin{equation}\label{GDE}
f_1 E + f_0 = 0,
\end{equation}
where $f_1 , f_0$ are functions of the coordinates, both explicitly and through
$V$ and its derivatives. 

If we want that the looked for invariant has an identically vanishing Poisson bracket with the Hamilton {\it regardless of the value of the energy} (in other words that $I_2$ should be a {\it strong} invariant), eq.(\ref{GDE}) must be satisfied for every value of the parameter $E$ so that the two equations
\begin{align}
\label{A1} 
f_1 &= 0, \\
\label{A2}
f_0 &= 0,
\end{align}
must separately be satisfied. The first of these equations is simply
\begin{equation}\label{Dxy}
D_{xy}= 0.
\end{equation}
As a consequence of this, the second, by a direct comparison, turns out to  coincide with the standard Darboux equation (\ref{darboux}). This time, the solution of equation (\ref{Dxy}), together with (\ref{WQIC1}) and (\ref{WQIC2}) for the leading order terms (now $D$ and $B$), gives
\begin{align}
D &= a (y^2 - x^2) + b y - d x + g, \label{LOWQIC1}\\
B &= - 2axy - b x - d y - e. \label{LOWQIC2}
\end{align}
Recalling (\ref{CK1}) and comparing with (\ref{LOQIC1}--\ref{LOQIC3}), we see that we have
arrived at a result completely equivalent to that of the standard Darboux-Whittaker approach. In fact, eq.(\ref{A2}) now coincides with eq.(\ref{darboux}) and, therefore, the same possible separable potentials in 2 dimensions can be found also following this alternative route. 

In our general presentation in section 4, we will see how this strategy reveals to be useful also in the case of higher degree invariants. While used mostly for its pedagogical role in the present instance, what is important to stress is the key role of the integrability condition in the form (\ref{GDE}): in the general case of a polynomial invariant of degree $M$, it turns out that eq.(\ref{GDE}) is a polynomial in $E$ of degree $M/2 $ for $M$ even and 
$(M+1)/2$ for $M$ odd.

In the case of invariance at fixed energy, the problem admits additional solutions. Now we have that eq.(\ref{Dxy}) is no more necessary and therefore 
(\ref{WQIC1}) and (\ref{WQIC2}) are unconstrained Cauchy-Riemann equations. Instead of (\ref{LOWQIC1}) and (\ref{LOWQIC2}), the solution is now given by an arbitrary analytic function: its real and imaginary parts respectively
provide the functions $D$ and $B$. Because of this feature, we will see in the following how it is more convenient to work with complex variables.

Summarizing, the above procedure shows that the strategy for finding weak invariants is well defined and points out where it must be constrained to get strong invariants. This objective being the most important for applications,
it may appear that, going this way, nothing is gained with respect to the usual direct approach. However, the procedure instead proves to be very effective in simplifying the system of equations resulting from implementing the direct approach. This simplification is actually one of the reasons why Hall's work is still useful. At the same time, since the approach based on the constrained invariant ansatz (\ref{weakI2}), complemented by a correct use of the general Darboux equation  (\ref{GDE}), provides the same results as the standard approach, we get a simple proof of the invalidity of Hall's criticism towards Whittaker. However, Hall is right in envisaging additional solutions to the problem: in the example we have just seen, it is conceivable to obtain, for {\it particular} values of $E$, solutions of eq.(\ref{GDE}) not included in those of \ref{A1} and \ref{A1}. In \cite{geom}, we have produced several
classes of solutions corresponding to $E=0$. 

The investigation of such {\it weakly integrable systems} (WIS) offers several additional issues to
study. We mention some of them:

\noindent
-- possible solutions defined for a continuous but finite range of energy values $E_1<E<E_2$;

\noindent
-- WIS which are actually SIS ({\it strongly integrable systems}) with a more general form for the second invariant (Examples: the Kepler problem, the Sarlet et al.\ case \cite{SLC85a});

\noindent
-- the main property of the integrable dynamics on the $E$ surface may help in understanding some of the main features in the non-integrable regime (see e.g.\ \cite{pr04}).


\section{The null Hamiltonian and time reparametrization}

As already put forward by Hietarinta (see \cite{hiet}, sect.7.2), canonical point
transformations generated by analytical functions preserve the form of a {\it null} Hamiltonian.
This invariance allows a straightforward way to reduce the set of equations to be solved in the search for a polynomial invariant. In the
present section we show how, in the case of 2-dimensional systems, the above transformations are
actually conformal transformations and are related with the time reparametrization of the dynamics.
These tools, exploited in the remaining part of the paper, were implicit in Hall's work.

In general the Hamiltonian itself has the form
   \begin{equation}  {  H} = T + V(q) = E ,
   \label{eq:origham}\end{equation}
where $T$ is a quadratic form in the momenta. The independent variable, let us say $t$, is often but
not always the time. For any given energy
$E$ of the system, to represent the dynamics, we can use the {\it null} Hamiltonian
\begin{equation}
{  H}_0 = {  H}-E ,
\end{equation}
 provided that we impose the constraint
   \begin{equation}  
{  H}_0 = 0 .
\end{equation}
For any such zero energy Hamiltonian we can reparametrize the system by
introducing a new time variable $\bar t$ defined by the relation
   \begin{equation}  dt = {  N}(p,q) d\bar t \ ,
\end{equation}
together with a redefined Hamiltonian
   \begin{equation}
       {\bar {  H}}_0 = {  N}(p,q) {  H}_0 = {  N} T + 
       {  N} (V-E) = 0 \ .
\end{equation}
The new Hamiltonian will then give the same equations of motion on the
constraint surface ${\bar {  H}}_0=0$.
We shall use the term {\it lapse function\/} for ${  N}(p,q)$ which defines
the independent variable gauge. This usage is borrowed from cosmological
applications where the lapse gives the rate of physical time change relative
to coordinate time. The lapse
function can be taken as any non-zero function on the phase space. 

It is simpler to work with complex, or ``null", coordinates, 
\begin{align}
z &= x+iy,       
\qquad p_z = p =  {\scriptstyle \frac12}  (p_x - i p_y), \label{eq:complex1}\\
\bar z &= x-iy,        
\qquad p_{\bar z} = \bar p =  {\scriptstyle \frac12}  (p_x + i p_y). \label{eq:complex2}
\end{align}
The null Hamiltonian can then be
written in the form
\begin{equation}\label{Hzero}
{  H}_0 = 2 p \bar p - G(z, \bar z) = 0,
\end{equation}
where $G$ is the function introduced in (\ref{G}). The equations of motion given by (\ref{Hzero}) are
\begin{align}
\frac{d z}{dt} & = 2 \bar p, \label{Hmotion1}\\ 
 \frac{d p}{dt} & = G_z = - V_z, \label{Hmotion2}
\end{align}
and corresponding complex conjugates. $G$ and $V$ are always assumed to be real functions. In the following we will spare to mention explicitly to the complex conjugates.

We use
a conformal transformation to standardize the frame and coordinate
representation of the invariant. To that end we introduce a new complex
coordinate $w$ by means of the transformation 
\begin{equation}\label{transf}
z =  F (w) .
\end{equation}
The conformal transformation 
\begin{equation}\label{zwtransf}
z \rightarrow w = X + i Y, 
\end{equation}
given by the holomorphic function
(\ref{transf}) determines the canonical point transformation
\begin{equation}
w =  F^{-1} (z), \quad
P = F' p, 
\end{equation}
so that (\ref{H}) transforms into the new null Hamiltonian
\begin{equation}\label{Hw}
\widetilde {  H}_0 = \frac{2 P \bar P - \widetilde G(w, \bar w)}
          {|F'|^2} = 0. 
\end{equation}
Let us introduce the ``standard" null Hamiltonian
\begin{equation}\label{HS}
{  H}_S = 2 P \bar P - \widetilde G(w, \bar w) = 0,
\end{equation}
that implies the use of the new ``time" $s$, such that
\begin{equation}
\frac{d}{d s} = |F'|^2 \frac{d}{d t}. 
\end{equation}
Equations of motion (\ref{Hmotion1}--\ref{Hmotion2}) becomes
\begin{align}
\frac{d w}{ds} & = 2 \bar P, \label{HSmotion1}\\ 
\frac{d P}{ds} & = \widetilde G_w. \label{HSmotion2}
\end{align}
At the same time, a conserved quantity stays conserved if transformed
between the two gauges (\ref{Hw}) and (\ref{HS}). In terms of real variables, (\ref{HS}) is given
by
\begin{equation}\label{HSR}
{  H}_S = \frac12 (P_X^2 + P_Y^2) - \widetilde G(X, Y) = 0.
\end{equation}

\section{Invariants of arbitrary degree}

With the choice of polynomial invariants, the time reflection symmetry of Hamiltonian
(\ref{H}) allows a further simplification in the procedure. In fact, it can been proved
(see Hietarinta \cite{hiet}, sect.2.3 and \cite{ny}) that the algebra of commuting functions with the
Hamiltonian, which is an even function with respect to time reflection, has ``good"
time reflection parity. Therefore, a polynomial second invariant can be either even
or odd polynomial in the momenta. We can therefore assume a phase-space function of the
following form:
\begin{equation}\label{invarb}
{  I}_M = \sum_{k=0}^{[M/2]} \sum_{j=0}^{M-2k}
p_x^j \, p_y^{M-2k-j} \, A_{(j,M-2k)} (x,y),
\end{equation}
where the functions $A_{(j,M-2k)} (x,y)$, not necessarily polynomials, have to be
determined. In the first summation, with $[M/2]$ we denote the greatest integer less
than $M/2$, so that if, e.g., $M=1$, $k$ takes only the value zero. 

\subsection{The direct approach in the Cartesian frame}
The simplest procedure (also called the {\it direct approach}) is now to compute the
Poisson brackets of
${  I}_M$ with
${  H}$, collect terms with various power of momenta end let vanish their respective
coefficients. Using at first the form (\ref{invarb}) in the usual Cartesian frame, we then get a
system of partial differential equations of the form:
\begin{equation}\label{invarbequa}
(j+1) A_{(j+1,k+1)} \partial_x V + (k+1-j) A_{(j,k+1)} \partial_y V =
\partial_x A_{(j-1,k-1)} + \partial_y A_{(j,k-1)}, 
\end{equation}
with $j = 0,...,k$ and $ k = M+1,M-1,M-3,...,0 \, {\rm or} \,1$ and it is implicit that $A_{(s,t)}=0$
if $s<0$ or $s>t$ and $t<0$ or $t>M$. In order to simplify formulas, in the present section we reintroduce the standard notation
$\partial$ to denote partial differentiation with respect to the variable in the subscript. The set
(\ref{invarbequa}) is a complicated set of PDEs, in general overdetermined: if $M$ is even we have
$(M+2)(M+4)/4$ equations for $(M+2)^2/4$ unknowns; if $M$ is odd we have
$(M+3)^2/4$ equations for $(M+1)(M+3)/4$ unknowns. Actually, in the general case, to these figures the potential $V$ enters as an additional unknown. 

\subsection{The direct approach in the null frame}
Introducing
the complex null frame (\ref{eq:complex1},\ref{eq:complex2}), we have the following generic expression of the invariant:
\begin{align}
{  I}_M = 2 {\rm Re}  \left\{ \sum_{k=0}^{M/2}
   {C_{2k} p^{2k}} \right\},  &\quad M \;\; {\rm even} \label{cinveven}\\[7pt]
{  I}_M = 2 {\rm Re}  \left\{ \sum_{k=0}^{(M-1)/2}
   {C_{2k+1} p^{2k+1}} \right\},  &\quad M \;\; {\rm odd} , \label{cinvodd}
\end{align}
where the complex functions $C_{2k}, k=0,...M/2$ or $C_{2k+1}, k=0,...(M-1)/2$ depend on $z,\bar z$
and, with the exclusion of the leading order coefficients $C_M$, implicitly on $E$. Their
explicit expression in terms of the coefficients of the invariant in Cartesian form are
\begin{equation}\label{cvaeven}
\begin{split} C_{2k}^{(M)} =
&\sum_{a=0}^{\frac{M}{2} - k} \left(\frac{G}{2}\right)^a \times \\
&\left\{ \sum_{\ell=0}^{2(a+k)} i^{2(a+k)-\ell}
\left[ \sum_{r=0}^a (-1)^r 
{\ell\choose{\ell-a+r}}{{2(a+k)-\ell}\choose{2(a+k)-\ell-r}}\right]
 A_{(\ell,2(a+k))} \right\}, \\ \end{split}
\end{equation}
when $ M $ is even and
\begin{equation}\label{cvaodd}
\begin{split} C_{2k+1}^{(M)} =
&\sum_{a=0}^{\frac{M-1}{2} - k} \left(\frac{G}{2}\right)^a  \times \\
&\left\{ \sum_{\ell=0}^{2(a+k)+1} i^{2(a+k)+1-\ell}
\left[ \sum_{r=0}^a (-1)^r 
{\ell\choose{\ell-a+r}}{{2(a+k)+1-\ell}\choose{2(a+k)+1-\ell-r}}\right]
 A_{(\ell,2(a+k)+1)} \right\},  \\ \end{split}
\end{equation}
when $ M $ is odd. In these expressions a superscript $(M)$ has been introduced in order to let
them apply in general: namely, $C_{a}^{(M)}$ denotes the $a-th$ coefficient of the invariant of
degree $M$. In the following this superscript will not be used unless it is strictly needed to
prevent confusion. As usual, in (\ref{cvaeven}) and (\ref{cvaodd}) 
\begin{equation}
{n\choose k} = \frac{n!}{(n-k)! \, k!}
\end{equation}
denotes the binomial coefficient.

The forms (\ref{cinveven}) and (\ref{cinvodd}) of the invariant do not contain cross terms, namely terms with powers of $p \bar p$, since they have been eliminated exploiting the constraint 
\begin{equation}\label{Hzeroconstraint}
2 p \bar p = G(z, \bar z) 
\end{equation}
dictated by (\ref{Hzero}). A simple check of this statement can be performed with the quadratic invariant with which we are familiar from Section 2. Eq.(\ref{cvaeven}) with $M=2$ give
\begin{align}
C_{2}^{(2)} &= A_{22} - A_{02} + i A_{12},\\
C_{0}^{(2)} &= A_{00} + (A_{02} + A_{22}) G.
\end{align}
Posing $A_{22} = A, \; A_{12} = B, \; A_{02} = C, \; A_{00} = K$ and $C_{2}^{(2)} = D + i B, \; C_{0}^{(2)} = {\widetilde K}$ we again find relations (\ref{CK1},\ref{CK2}) and see that the expression (\ref{weakI2}) of the invariant is equivalent to
\begin{equation}
{  I}_2 (p, \bar p, z, \bar z) = 
 C_{2}^{(2)} p^2 + {\bar C}_{2}^{(2)} {\bar p}^{2}+ {\widetilde K}. 
\end{equation}

The use of the energy constraint leads to the maximal reduction in the number of equations ensuing from the direct approach. However, in the case of integrability at arbitrary energy, to this reduced number one must add the integrability conditions which results in a set of equations generalizing (\ref{A1}). The constraint (\ref{Hzeroconstraint}) can also be used to ``homogeneize" the polynomial invariant. In this form, used in the geometric
framework in connection with the geodesic flow over a Riemannian manifold, the coefficients of the invariant give rise to a symmetric M-rank tensor, known as a {\it Killing tensor}.

The commutation relation of the functions (\ref{cinveven}) or (\ref{cinvodd}) with the null Hamiltonian (\ref{Hzero}) gives the system of equations 
\begin{equation}\label{cinv2}
\partial_{\bar z} C_{k-1} 
+ \tfrac12 (\partial_z C_{k+1}) G 
+ \tfrac12 (k+1) C_{k+1} \partial_z G =0,
\quad k = 0, 1, \dots, M,
\end{equation}
where it is 
implicitly assumed that $C_j = 0$ for $j<0$ and for $j>M$. The set of 
equation (\ref{cinv2}) must be supplemented by the closure equations 
\begin{equation}\label{cinv1} 
\partial_{\bar z} C_M = 0 \end{equation} 
and 
\begin{equation}\label{cinvclo1}
\Re \{\partial_z (C_1 G) \}=0.
\end{equation}
Clearly, this last condition must be stated only in the case of odd degree, since for $M$ even, a closure condition of the form
\begin{equation}\label{cinvclo2}
\Im \{C_0 \}=0
\end{equation}
is implicit since $C_{0}$ is real by definition.

In both cases, using the null complex coordinates, we have $M+2$ real equations for $M+1$ real unknowns: the substantial reduction of the number of independent equations with respect to that in the Cartesian frame was already remarked by Hietarinta (see \cite{hiet}, sect.7.5). However, he thought that reintroducing the explicit dependence on energy would reestablish the original number, whereas, as we will see, the gain in saving equation still remains when going to the Cartesian frame. 

\subsection{Solving the equations for the Mth-degree invariant}
At present a general solution of system (\ref{cinv2}) is still lacking. In  \cite{geom,max,kprs} we found the general solutions for $1 \le M \le 4$. In the present subsection we illustrate the aspect of the procedure which are common to all Mth-degree invariant. The first steps are essentially the same as in the original paper by Hall. 

Looking at the system above, we see that eq.(\ref{cinv1}) is readily solved:
\begin{equation}\label{cm}
C_M = C_M (z),
\end{equation}
that is $C_M$ is an arbitrary holomorphic function. The first important result we get is therefore that the leading order coefficient of a polynomial invariant is given by an arbitrary holomorphic function. This fact is already known in the case of homogeneous polynomial invariants (Kolokoltsov \cite{kolokoltsov:inv}, Kozlov \cite{kozlov}) and was obtained by Birkhoff (\cite{birkhoff}, chap. 2) for $M\le2$. We remark that, in agreement with notations in (\ref{cvaeven}) and (\ref{cvaodd}), here we are referring to $C_M^{(M)}$, that is the leading order coefficient of the polynomial invariant of degree $M$ with various values of $M$. Therefore, in order to avoid confusion with lower order coefficients in a polynomial with a given $M$ and also to conform with the notation in previous works, in the following we will denote the leading order coefficient with $S_M$.

In order to attack the remaining equations in (\ref{cinv2}), we may further simplify them by performing a coordinate transformation. This is a conformal
transformation of the form (\ref{transf}), where the generating function $F(w)$
is chosen such that
\begin{equation}\label{FM}
F'(w(z)) = {S_M}^{1/M}.
\end{equation}
In this case, we see that the first equation of the chain (\ref{cinv2}) can be rewritten as
\begin{equation}\label{cinv22}
{S_M}^{{\frac{2}{M}} - 1} \partial_{\bar w} C_{M-2} +{\scriptstyle \frac{M}2} \partial_w {\widetilde G} = 0,
\end{equation}
where, in agreement with (\ref{Hw}), the new ``conformal" potential
\begin{equation}
\widetilde G(w, \bar w) = |F'(w)|^2 G=(S_M \bar S_M)^{1/M} G,
\end{equation}
has been introduced. Defining the function
\begin{equation}\label{tildeR}
\widetilde C_{M-2}(w, \bar w) = S_M^{\frac{2-M}{M}} C_{M-2},
\end{equation}
eq.(\ref{cinv22}) becomes:
\begin{equation}\label{cinv222}
\partial_{\bar w} \widetilde C_{M-2} + {\scriptstyle \frac{M}2} \partial_w {\widetilde G} = 0
\end{equation}
and, defining in an analogous way,
\begin{equation}
\widetilde C_{2k}(w, \bar w) = S_{M}^{-\frac{2k}{M}} C_{2k}, \quad k = 0,1,\dots,
[N/2],
\end{equation}
the rest of the system becomes
\begin{equation}\label{finale}
\partial_{\bar w} \widetilde C_{j-2} + 
\frac{1} {2 \widetilde G^{j-1}} \partial_w (\widetilde C_j \widetilde G^j)  
=0,
\end{equation}
with the ranges $j = 4,5,\dots,N-2$ if $N$ is even and $j = 3,4,\dots,N-2$ if $N$ is odd.

\subsection{The K\"ahler potential}
Eq.(\ref{cinv22}), or better its transformed form (\ref{cinv222}), plays a special role. In view of the fact that the new potential $\widetilde G$ is still a real function, (\ref{cinv222}) can be solved introducing a real function ${\cal  K}(w, \bar w)$ such that
\begin{align}
\widetilde G &= \partial^2_{w \bar w}{\cal  K}, \label{Kpot} \\
\widetilde C_{M-2} &= -{\scriptstyle \frac{M}2}\partial^2_{w w}{\cal  K}. \label{KC}
\end{align}
The first equation says that the Jacobi potential is the Laplacian of the function ${\cal  K}(w, \bar w)$. Since, in the geometric picture, $G$ is the conformal factor of the metric element, using ${\cal  K}$ it gives rise to a Hermitian form and therefore it is referred to as the {\it K\"ahler
potential}. Without loss of generality, in (\cite{kprs}, sect. IV) it has been shown that the K\"ahler potential can be assumed to be of the form
\begin{equation}\label{KEP}
{\cal  K} = E [ F(w) \bar F(\bar w) + 2 \Re \{ \Lambda(z(w))\}] - \Psi,
\end{equation}
where $\Lambda$ is an arbitrary holomorphic function independent of $E$ and the real {\it
prepotential} $\Psi$ is such that
\begin{equation}
\partial^2_{w \bar w}\Psi = 
F' \bar F' \partial^2_{z \bar z} \Psi = |F'|^2 V .
\end{equation}
In the case of invariants up to the fourth degree, $M \le 4$, we see that the closure
equations (\ref{cinvclo1}) and (\ref{cinvclo2}) are themselves expressed only in terms of the K\"ahler potential. In these cases, to get a
complete solution of the problem, it remains {\it only} to solve an integrability condition for ${\cal  K}(w, \bar w)$. 

In the cases of linear and quadratic invariants,
the integrability condition is linear and we have the general solution depending respectively on one
and two arbitrary real functions (see \cite{geom}). In the higher degree cases, the integrability conditions
are nonlinear and a general solution is lacking: moreover, in general, they give rise to an
overdetermined system for the unknowns $\Lambda$ and
$\Psi$. This is consistent with the fact that only isolated cases of integrable Hamiltonian systems
with invariant of degree higher than two are known. Failing to work in full generality, a useful
approach is to make an ansatz for one of the unknown (e.g. $\Lambda$) and solve for $\Psi$. In this way several new integrable and superintegrable systems have been found with a cubic (\cite{max}) and a quartic ({\cite{kprs}) second invariant. 

\subsection{The leading order term of strong invariants}
Looking for a strong invariant, so that the integrability condition must be solved for arbitrary values of $E$, it turns out that the function $C_M$
can no longer be arbitrary. In our previous works we have shown its polynomial structure with degree equal to $M$ for $M\le4$ and guessed that this is the generic behavior for any value of $M$. In this section we prove that this conjecture is true by explicitly computing the function with the
aid of the already known solution in the Cartesian case. 

In \cite{geom} for the linear and quadratic cases we have found respectively the conditions
\begin{align}
\Re  \{ S_1'(z) \} &= 0, \label{S1} \\
\Im  \{ S_2''(z) \} &= 0. \label{S2} 
\end{align}
In \cite{max}, for the cubic case, the condition
\begin{equation}\label{S3}
\Re  \left\{ \frac{d^{3}}{dz^{3}}  S_3\right\} = 0
\end{equation}
was found, whereas in {{\cite{kprs}}, for the quartic case, the condition
\begin{equation}\label{S4}
\Im  \left\{ \frac{d^{4}}{dz^{4}}  S_4 \right\} = 0
\end{equation}
was found. Therefore we can guess the general conditions
\begin{align}
\Re  \left\{ \frac{d^{M}} {dz^{M}}  S_M \right\}  &= 0, \;\; M \;\; {\rm odd},  \label{SModd} \\
\Im  \left\{ \frac{d^{M}} {dz^{M}}  S_M \right\}  &= 0, \;\; M \;\; {\rm even}. \label{SMeven} 
\end{align}
Conditions from (\ref{S1}) to (\ref{S4}) have been obtained working on the integrability condition for the K\"ahler potential ${\cal  K}$. Since it is not possible to write an explicit expression for the integrability condition at arbitrary $M$, we prove (\ref{SModd}) and (\ref{SMeven}) by explicitly calculating them. The easiest way to proceed is to exploit the solution in Cartesian coordinates and after that performing the transformation to complex coordinates: the reason for not working directly in the complex frame is that, in (\ref{cinveven}) and (\ref{cinvodd}), part of information on the structure of $S_M$ is hidden in the lower order coefficients through the energy constraint: this information could be recovered only having the solution of the complete system of equations for $C_k$ up to the integrability for $C_0$.

Comparing (\ref{cinveven}) and (\ref{cinvodd}) with (\ref{invarb}) we get the relation 
\begin{equation}\label{Scart}
S_M (x + i y) \equiv C_M = \sum_{m=0}^M i^{M-m} A_{(m,M)} (x,y).
\end{equation}
As it is natural, the leading order complex coefficient depends only on leading order Cartesian coefficients. Considering (\ref{invarbequa}) with $k = M+1$, we get the equation for the leading order Cartesian coefficients, namely
\begin{equation}
\partial_x A_{(j-1,M)} + \partial_y A_{(j,M)} = 0,
\quad j = 0,1,...,M+1,
\end{equation}
whose solution can be immediately found (see Hietarinta \cite{hiet}, sect.3.0):
\begin{equation}\label{ALOT}
A_{(m,M)} = \sum_{k=0}^m \sum_{j=0}^{M-m} (-1)^k
{{(j+k)!}\over{j!k!}} a_{(j+k,m-k,M)} x^j y^k
\end{equation}
where $a_{(j+k,m-k,M)}$ are {\it real} integration constants.

From the relation
\begin{equation}
\partial_z = {\scriptstyle \frac12} (\partial_x - i \partial_y)
\end{equation}
we have the expression for the $r-th$ derivative 
\begin{equation}\label{pzr}
\partial_z^r = {1 \over {2^r}} \sum_{s=0}^r {{r!} \over {(r-s)!s!}}
(-i)^s \partial_x^{r-s} \partial_y^s.
\end{equation}
In order to compute the operator (\ref{pzr}) we need the following intermediate result
\begin{equation}\label{intr1}
\partial_x^{r-s} \partial_y^s x^jy^k= 
{{j!} \over {(j-r+s)!}}{{k!} \over {(k-s)!}}x^{j-r+s} y^{k-s}, 
\;\; {\rm for} \;\; k \ge s \;\; {\rm and} \;\; j \ge r-s 
\end{equation}
whereas the result is zero if $ k < s $ or $ j < r-s $. Using it, 
the action of the operator (\ref{pzr}) on the monomial $x^jy^k$ is
\begin{equation}\label{intr2}
\partial_z^r x^jy^k = {1 \over {2^r}} \sum_{s=0}^r (-i)^s 
{{r!} \over {(r-s)!s!}} {{j!k!} \over {(j-r+s)!(k-s)!}}
 x^{j-r+s} y^{k-s}.
\end{equation}
Actually, we are interested only in the in the action of $\partial_z^M$: in this case, since each
derivative corresponds to decreasing the degree of the monomial by one, only the highest degree
terms, with $j+k=M$, survive the action of $\partial_z^M$. Applying (\ref{intr2}) we get:
\begin{equation}\label{intr3}
\partial_z^M x^{M-k}y^k = {1 \over {2^M}} \sum_{s=0}^M (-i)^s 
{{M!} \over {(M-s)!s!}} {{k! (M-k)!} \over {(s-k)!(k-s)!}}
x^{s-k} y^{k-s}.
\end{equation}
Analogously to those in (\ref{intr1}), we have that both conditions $s-k \ge 0$ and $k-s \ge
0$ must be satisfied. They implies just $s=k$ so that (\ref{intr3}) turns out to be simply
\begin{equation}\label{intr4}
\partial_z^M x^{M-k}y^k = {1 \over {2^M}} (-i)^k
{{M!} \over {(M-k)!k!}} k! (M-k)! = (-i)^k {1 \over {2^M}} M!.
\end{equation}
The same result holds exchanging $k$ with $M-k$:
\begin{equation}\label{intr5}
\partial_z^M x^ky^{M-k} = (-i)^k {1 \over {2^M}} M!.
\end{equation}
Let us now denote with $\hat{A}_{(m,M)}$ the highest degree part of $A_{(m,M)}$: 
\begin{equation}\label{intr6}
\hat{A}_{(m,M)} = (-1)^m
{{M!}\over{(M-m)!m!}} a_{(M,0,M)} x^{M-m} y^m.
\end{equation}
Using (\ref{intr4}) we then have
\begin{equation}
\partial_z^M A_{(m,M)} = \partial_z^M \hat{A}_{(m,M)}  = 
{{i^m} \over {2^M}} {{M!}^2\over{(M-m)!m!}} a_{(M,0,M)}.
\end{equation}
Remembering (\ref{Scart}) we finally get
\begin{align}
\partial_z^M S_M &= \sum_{m=0}^M i^{M-m} \partial_z^M A_{(m,M)} \\
&= \sum_{m=0}^M i^{M-m} {{i^m} \over {2^M}} {{M!}^2\over{(M-m)!m!}} a_{(M,0,M)} \\
&= {{i^m} \over {2^M}} M! a_{(M,0,M)} \sum_{m=0}^M {{M!}\over{(M-m)!m!}} \\
&= i^m M! a_{(M,0,M)}
\end{align}
which is just what we wanted to prove since it is equivalent to (\ref{SModd}) and (\ref{SMeven}).

\section{Examples}

We illustrate the applications of the general approach described so far with a selection of results, many of which are new. In particular, we provide some       weakly integrable systems with linear and quadratic invariants and a recipe for higher order examples. In order to compactify formulas,  from hereinafter with the subscript 
we denote the partial derivative with respect to the corresponding 
variable (except when used to denote a component of momentum).

\subsection{Weakly integrable systems with linear invariants}
The ansatz is 
\begin{equation}
\label{line}  {  I}_1 = S p + {\bar S} {\bar p},  
\end{equation}  
where for simplicity we have suppressed the subscript in the $S$ function. The system of equations ensuing 
from the conservation condition is the following:  
\begin{align}  
     S_{\bar z} &= 0, \label{ic11}\\  
 \Re \{(G S)_z\} &= 0. \label{ic12} 
\end{align}
Equation (\ref{ic11}) agrees with 
(\ref{cm}), confirming that $S$ can be an arbitrary analytic 
function, 
\begin{equation}\label{SS} 
S = S (z). 
\end{equation} 
According to (\ref{FM}), the conformal transformation is given by 
\begin{equation}\label{cF1} 
\frac{dz}{dw} 
= F'(w) \equiv S(z(w)) \end{equation} 
or equivalently 
\begin{equation}\label{icFi1} 
w = X + i Y = \int \frac{dz}{S(z)}. 
\end{equation}
Introducing the ``conformal'' 
potential  
\begin{equation}\label{cpot} \widetilde G = |F'|^2 G = |S|^2 G = S \bar S 
G, \end{equation} 
eq.(\ref{ic12}) reduces to 
\begin{equation}\label{intc1} 
\Re \{{\widetilde G}_w\}=0, 
\end{equation} 
which is readily solved in 
\begin{equation}
\label{igs} \widetilde G = g (Y), 
\end{equation} 
where 
$g$ is an arbitrary real function and, according to the definition of 
the coordinate transformation, $Y$ is the imaginary part of $w$. In this coordinates, the invariant assumes the `normal form'
\begin{equation}
\label{linew}   I_1 = P + {\bar P} 
\end{equation} 
or simply
\begin{equation}
\label{linewc}   I_1 = P_{X}. 
\end{equation} 
Equation (\ref{ic12}), in view of (\ref{SS}) and recalling the 
definition of $G$, can be rewritten as 
\begin{equation}\label{G1} \Re \{S'( E-V) - S V_z\}=0. \end{equation} 
We recognize the structure of eq.(\ref{GDE}), where now
\begin{equation}\label{iS1} 
f_1 = \Re \{S'(z)\}=0 \end{equation}  
and 
\begin{equation}\label{V1}  f_0 = \Re \{(S V)_z\}=0
 \end{equation} 
and we see that (\ref{iS1}) coincides with (\ref{S1}) and solution (\ref{igs}) now applies to $V$, so that we have
\begin{equation} \label{V0} 
V = \frac{g(Y(x,y))}{|S|^2}, \end{equation} 
with $g$ arbitrary. The general form of the second invariant in the original real coordinates is 
\begin{equation}\label{I1} 
 I_1 = \Re \{S\} p_{x} + \Im \{S\} p_{y}. \end{equation}

We may provide two interesting classes of weakly 
integrable systems admitting linear invariants. The first is 
obtained by the simple observation that, if we choose the level 
surface $ E=0$, it is no longer necessary that condition (\ref{iS1}) be 
satisfied. {\it Any} analytic function $S=S(z)$ provides a solution 
through the corresponding conformal transformation. If $Y$, as above, 
denotes the new coordinate 
\begin{equation} Y = 
{\rm Im} \left\{\int \frac{dz}{S(z)}\right\}, \end{equation} 
then the solution is given by the pair (\ref{V0}--\ref{I1}), provided we consider only motions at the energy level $ E=0$.

The second 
class of weakly integrable systems is obtained with the following 
trick. Let us consider the analytic function $f(z)$ and consider then 
the conformal transformation (\ref{cF1}) with $S(z)$ given by 
\begin{equation}\label{sweak} 
S = \frac{1}{a + f'(z)}, 
\end{equation} 
with $a$ constant. 
Recalling definition (\ref{cpot}), let us consider the ``flat" 
conformal potential $\widetilde G = 1$. In this case, relation 
(\ref{cpot}), using (\ref{sweak}) gives 
\begin{equation} G = E-V = 
\frac{1}{|S|^2} = a^2 + a (f'+ \bar f') + |f'|^2. 
\end{equation} 
We can 
therefore interpret $a^2$ as the fixed value of the energy constant 
and get as a consequence the family of potentials  
\begin{equation} 
V(z, \bar z;  E) = - \sqrt{ E} (f'+ \bar f') - |f'|^2.  
\end{equation} 
To complete the 
solution, we have to write explicitly the coordinate transformation 
generated by (\ref{sweak}), that is 
\begin{equation} 
w = \int \frac{dz}{S(z)} = a z + f(z). \end{equation} 
The invariant is still of the form (\ref{I1}).

\subsection{Weakly integrable systems with quadratic invariants}
The ansatz is 
\begin{equation}
\label{quad}   I_2 = S p^{2} + {\bar S} {\bar p}^{2} + \widetilde K.  
\end{equation}  
The system of equations ensuing 
from the conservation condition is the following:  
\begin{align}
 S_{\bar z} &= 0, \label{ic21}\\  
\widetilde K_{\bar z} + S G_z + {\scriptstyle \frac12} S' G &= 0 \label{ic22}. 
\end{align}  
Equation (\ref{ic21}) is already familiar. Since $\widetilde K$ is real, equation (\ref{ic22}) has the following integrability condition 
\begin{equation}\label{ic23}
\Im \{S''G + 3 S' G_{z} + 2 S G_{zz}\} = 0.
\end{equation}
However, as above, we can directly simplify eq.(\ref{ic22}). According to (\ref{FM}), the conformal transformation is now given by 
\begin{equation}\label{cF2} 
\frac{dz}{dw} 
= F'(w) \equiv \sqrt{S(z(w))} \end{equation} 
or equivalently 
\begin{equation}\label{icFi2} 
w = \int \frac{dz}{\sqrt{S(z)}}. 
\end{equation}
With the conformal potential  
\begin{equation}\label{cpot2} 
\widetilde G = |F'|^2 G = |S| G = \sqrt{S \bar S} G,
\end{equation} 
eq.(\ref{ic22}) reduces to 
\begin{equation}\label{cKG2} 
\widetilde K_{\bar w} + \widetilde G_w=0, 
\end{equation} 
that is the first example of the set (\ref{cinv222}). Its integrability condition is 
\begin{equation}\label{intc2}
\Im \{\widetilde G_{ww}\} = 0
\end{equation}
which is readily solved by 
\begin{equation}\label{igsw} \widetilde G = {\widetilde A}(w+\bar w) + {\widetilde B} (w - \bar w), \end{equation} 
where $ {\widetilde A}$ and $ {\widetilde B}$ are arbitrary real functions. Putting this solution in (\ref{ic22}) gives
\begin{equation}\label{cK2} 
\widetilde K= {\widetilde B} -  {\widetilde A}.
\end{equation} 
Comparing 
with (\ref{cpot2}) and using real ``separating'' coordinates, 
\begin{equation}
X = \Re\{w\}, \;\; Y = \Im\{w\},\end{equation} 
the solution for the original function $G$ is 
then 
\begin{equation}\label{igs2} 
G = \frac{{\widetilde A}(X;  E) + {\widetilde B} (Y;  E)}{|S|}. 
\end{equation} 
The invariant takes the `normal form'
\begin{equation}
\label{quadw}   I_2 = P^{2} + {\bar P}^{2} + {\widetilde B} -  {\widetilde A}
                                    = \frac12 (P_{X}^{2} - P_{Y}^{2}) + {\widetilde B} -  {\widetilde A}
\end{equation} 
or, in the original real coordinates,
\begin{equation}
\label{quadwc} I_2 =  \frac12 {\rm Re} (S) (p_x{}^2 - p_y{}^2) + {\rm Im} (S) p_x p_y
+ {\widetilde B (Y(x,y))} -  {\widetilde A (X(x,y))}. 
\end{equation} 
Equation (\ref{intc2}) can be rewritten as 
\begin{equation}\label{G2} 
\Im \{S'' ( E-V) - 3 S' V_{z} - 2 S V_{zz}\} = 0. \end{equation} 
We again recognize the structure of eq.(\ref{GDE}), where now
\begin{equation}\label{iS2} 
f_1 = \Im \{S''(z)\}=0 \end{equation}  
and 
\begin{equation}\label{V2} 
 f_0 = \Im \{S''V + 3 S' V_{z} + 2 S V_{zz}\} = 0
 \end{equation} 
and we see that (\ref{iS2}) coincides with (\ref{S2}). It can 
be proven (see \cite{pr04}, section 2) that this condition, in addition to warrant strong integrability,  
is also necessary and sufficient in order that the 
conformal factor $|S|$ can be written as a {\it sum} of two functions of 
$X$ and $Y$,
say
\begin{equation}\label{sab}
|S| = S_{X}(X) + S_{Y}(Y).\end{equation} 
Since from (\ref{V2}) we have that solution (\ref{igs2}) now applies to $V$, we get
\begin{equation} \label{VR} 
V =  \frac{{A}(X) + {B} (Y)}{|S|}. 
\end{equation}  
with suitable $A$ and $B$. In this case, $X$ and $Y$ are properly referred to as {\it separable} coordinates. From (\ref{igs2}) and (\ref{sab}), we deduce the relations
\begin{align}
 {\widetilde A} (X;  E) &=  E S_{X} - A(X),\\
 {\widetilde B} (Y;  E) &=  E S_{Y} - B(Y),
\end{align}
so that, eliminating the energy parameter through the Hamiltonian 
\begin{equation}\label{HSRE}
 E = {  H} = 
 \frac{{\scriptstyle \frac12} (P_X^2 + P_Y^2) + A + B}
 {S_{X} + S_{Y}},
\end{equation}
  the general form of the strong quadratic invariant is 
\begin{equation} 
{  I}_2 = \frac{S_{Y} (P_{X}^{2} + 2 A) - S_{X} (P_{Y}^{2} +  2 B)}{S_{X} + S_{Y}}.
\end{equation}

We mention here some interesting examples of weak integrability with quadratic invariants. Consider first the polynomial function $S(z) = i z^2$. This is the simplest
polynomial which gives a potential which is not automatically integrable at
arbitrary energy. The corresponding conformal transformation is given by 
\begin{equation}
w = 2^{-1/2} (1-i) \ln z
\end{equation} 
or in terms of the real variables using polar coordinates
\begin{align}
      X &= \frac1{\sqrt2} (\theta + \ln r) \\
      Y &= \frac1{\sqrt2} (\theta - \ln r).
\end{align}
From the relation (\ref{igs2}) it then follows that the potential
given by
\begin{equation}
      V = r^{-2} [ A(r e^{\theta}) + B(r e^{-\theta}) ] ,
\end{equation}
is integrable at zero energy for arbitrary functions $A$ and $B$. In \cite{pr04}, the class of systems with
\begin{align}
      A(X) &= \frac{1}{2} (C - \sin {\sqrt2} X) ,\\
      B(Y) &= \frac{1}{2} (C - \sin {\sqrt2} Y) ,
\end{align}
where $C$ is a real constant, has been investigated. Using polar coordinates, the explicit form of the potential is
\begin{equation}\label{iiz2pot}
       V(r, \theta) = \frac{C - \sin \theta \cos (\ln r)}{r^2} 
\end{equation}
and that of the second invariant is
\begin{equation}\label{secondiz2}
 {  I}_2 (p_r, p_{\theta}, r, \theta) =
      r p_r p_{\theta} -\cos \theta \sin (\ln r) 
 \end{equation}
where
\begin{align}
      r p_r &= x p_x + y p_y \\
 p_{\theta} &= x p_y - y p_x .
\end{align}
A natural question one can ask is if the system is still integrable at arbitrary energy values. As it is well known, it is possible to answer this question by performing careful numerical investigations, but, on purely analytical grounds, the problem is in general quite difficult. In \cite{pr04}, this potential has been proven to be non-integrable at values of the energy $-\epsilon_1 < E < \epsilon_2$, with small enough 
$\epsilon_{1,2}$, for every value of $C$ in the interval  $(0,1]$. The proof has been obtained by applying the Poincar\'e non-existence theorem of additional invariants (see, e.g. \cite{Poincare,akn}), working in the transformed coordinates $X,Y$, taking as unperturbed part of the Hamiltonian the integrable system at $E=0$ and as a small perturbation the term associated to the conformal factor.

A second example is that given by a generating function of the form
 \begin{equation}\label{z-2}
 S(z) = z^{-2} .
 \end{equation}
 The conformal transformation can then be written in terms of the real
 variables as
\begin{align}
 X &= {\scriptstyle \frac12} r^2 \cos 2 \theta
           = {\scriptstyle \frac12}(x^2-y^2) ,\\
 Y &= {\scriptstyle \frac12} r^2 \sin 2 \theta = xy .
\end{align}      
 The corresponding potential is
 \begin{equation}
        V(X,Y) = r^2 [ A(X) + B(Y) ] .
 \end{equation}
 since the conformal factor is
 \begin{equation}\label{confact2}
 |S(X,Y)| = \frac{1}{2\sqrt{X^2 + Y^2}} = \frac{1}{r^2} .
 \end{equation}
 Choosing for example
 \begin{align}\label{z-2AB}
       A(X) &= 4 X^2 ,\\
       B(Y) &= (2 + a) Y^2 + b,
 \end{align}
 where $a$ is a constant such that $0 \le a \le 2$ and $b > 0$, we get the
 family of potentials
 \begin{equation}\label{eq:pot2}
        V(x,y) = r^2 ( x^4 + y^4 + a x^2 y^2 + b ) .
 \end{equation}
 This potential has a relative maximum at the origin where $V(0,0) = 0$, absolute minima placed
 symmetrically in the four quadrants around the origin and grows as $r^6$ for
 $r \rightarrow \infty$.  It allows bound motion for all the admissible values
 of the energy.  For $a=2$ it is rotationally symmetric and therefore it is
 ``superintegrable" at zero energy. The second invariant is
\begin{equation}
{  I}_2 (p_x, p_y, x, y) =
               \frac{x^2 - y^2}{2r^4} (p_x^2 - p_y^2) -
               \frac{2 x y}{r^4} p_x p_y + 
               x^{4} + y^{4} - (4+a) x^{2} y^{2}.
 \end{equation} 
What is remarkable in this case is that already at energies slightly below or above the integrable level $E = 0$, the dynamics is strongly {\it chaotic}. This shows that identifying weakly integrable systems does not necessarily provide a {\it nearly} integrable system.

 \subsection{Weakly integrable systems with higher-order invariants}
 Exploiting the results obtained in \cite{max,kprs}, several families of weakly integrable systems admitting cubic and quartic invariants can be obtained. Both cases can be described with a similar approach. The invariants have the form
\begin{align}
I_3 &= 2 {\rm Re} \{S_{3} p^{3} + R_{3} p\},\\
I_4 &= 2 {\rm Re} \{S_{4} p^{4} + R_{4} p^{2}\} + C_{0}.
\end{align}
The conformal transformations (\ref{FM}) respectively give the normal forms
\begin{align}
I_3 &= 2 {\rm Re} \{P^{3} + {\widetilde R}_{3} P\},\\
I_4 &= 2 {\rm Re} \{P^{4} + {\widetilde R}_{4} P^{2}\} + C_{0},
\end{align}
with
\begin{equation}
{\widetilde R}_{M} = S_M^{{2-M}\over M} R_{M}, \;\;\; M =3,4,
\end{equation}
in agreement with definition (\ref{tildeR}). Introducing the K\"ahler potential, eq.(\ref{KC}) provides the solutions for ${\widetilde R}_{M}$:
\begin{equation}
{\widetilde R}_{M} = -{\scriptstyle \frac{M}2} {\cal  K}_{w w}. \label{KCM34}
\end{equation}
In the cubic case, the closure equation (\ref{cinvclo1}) expressed in terms of the K\"ahler potential gives
\begin{equation}\label{intc3}
\Re \{({ \cal K}_{w w} {\cal  K}_{w \bar w})_{w}\} = 0.
\end{equation} 
In the quartic case, the closure equation is that determined by eq.(\ref{finale}) that, with $j=2$ and the notations adopted here, gives
\begin{equation}\label{finale4}
 (C_{0})_{\bar w} = 
- {1 \over {2 \widetilde G}} (\widetilde R_{4} \widetilde G^2)_w.  \end{equation}
The reality of $C_{0}$ requires the
integrability condition 
\begin{equation}\label{intc4}
\Im \{({\cal  K}_{w w w} {\cal  K}_{w \bar w} + 2 
        {\cal  K}_{w w \bar w} {\cal  K}_{w w})_{w}\} = 0.
\end{equation} 
Equations (\ref{intc3},\ref{intc4}) illustrate the important difference between the previous linear and quadratic cases ($M=1,2$) and the higher order ($M>2$) cases. In the linear (see eq.(\ref{intc1})) and the quadratic case (see eq.(\ref{intc2})), the integrability conditions are {\it linear} differential equations: their solution is in terms of one or two arbitrary functions respectively. Now, equations (\ref{intc3}) and (\ref{intc4}) are both {\it nonlinear} partial differential equations: they appear very complicated and in practice impossible to solve in full generality. Only isolated systems can be identified. Using the general expression (\ref{KEP}), we get in both $M=3$ and $M=4$ cases a {\it quadratic} equation of the form
\begin{equation}\label{iS34} 
f_2 E^{2} + f_1 E + f_0 = 0,
\end{equation}
where the $f_k, \ k=0,1,2$ depend on $S$, $\Lambda$, $\Psi$ and their derivatives. Looking for strong invariants, $S$ is determined as above (cfr. eq.(\ref{S3}) and eq.(\ref{S4})) and several solutions can be obtained by suitable assumptions on $\Lambda$ and $\Psi$ \cite{max,kprs}. 

Looking for weak invariants, a large class of solutions can be obtained in the simplest eventuality $E=0$. In fact, in this case the only condition we have to satisfy is $f_0 = 0$ and it can be proven that this equation is a nonlinear PDE in the variable $\Psi$ only. In fact, we get respectively
\begin{equation}\label{psi3}
\Re \{({\Psi}_{w w} {\Psi}_{w \bar w})_{w}\} = 0, \ \ M=3
\end{equation} 
and
\begin{equation}\label{psi4}
\Im \{({\Psi}_{w w w} {\Psi}_{w \bar w} + 2 
        {\Psi}_{w w \bar w} {\Psi}_{w w})_{w}\} = 0, \ \ M = 4.
        \end{equation}
Now, {\it any} solution of these equations in whatever coordinate system can be used to generate a new weakly integrable system (at $E=0$) using an arbitrary conformal transformation not included in the families (\ref{S3}) and (\ref{S4}).        

\section{Conclusions}

Much of the structure and ideas presented in this work has come from the geometric formulation using the Jacobi metric as described in \cite{geom} and further developed in \cite{max} and \cite{kprs}. One of the important results of those earlier works was the unification of the notions of weak and strong integrability in a common framework. Another was the identification of the crucial role of the conformal transformations for analyzing integrability in two dimensions. However, while the geometric picture has been very useful in these and other respects, it is clear from the present work that an approach which is more closely related to the more traditional Hamiltonian picture can also be very effective. 

The conformal transformations have provided the clue for finding explicit conditions for the leading order term of strong invariants of arbitrary degree. The conditions, as given in equations (\ref{SModd}) and (\ref{SMeven}), are natural generalizations of the conditions for low order invariants. The examples given in section 5 provide an illustration of the apparent ease by which it is possible to construct interesting weak invariants. We hope that this work can serve as a starting point for improving the understanding of higher order invariants of both the strong and the weak type.


\end{document}